\documentclass[10pt,a4paper]{article}
\usepackage{graphicx}
\title{%
{\vspace{-3cm} \normalsize
     \hfill\parbox{38mm}{MS-TP-04-10 \\
                         cond-mat/0405673}  }\\[20mm]
 Profile and width of rough interfaces}
\author{Melanie M\"uller and Gernot M\"unster%
\thanks{Institut f\"ur Theoretische Physik,
        Universit\"at M\"unster,
        Wilhelm-Klemm-Str.~9, D-48149 M\"unster, Germany;
        e-mail: einalem@uni-muenster.de, munsteg@uni-muenster.de}}
\date{May 27, 2004\\(revised: October 27, 2004)}
\begin{document}
\maketitle

\begin{abstract}
In the context of Landau theory and its field theoretical refinements,
interfaces between coexisting phases are described by intrinsic
profiles. These intrinsic interface profiles, however, are neither
directly accessible by experiment nor by computer simulation as they
are broadened by long-wavelength capillary waves.  In this paper we
study the separation of the small scale intrinsic structure from the
large scale capillary wave fluctuations in the Monte Carlo simulated
three-dimensional Ising model. To this purpose, a blocking procedure is
applied, using the block size as a variable cutoff, and a
translationally invariant method to determine the interface position of
strongly fluctuating profiles on small length scales is introduced.
While the capillary wave picture is confirmed on large length scales
and its limit of validity is estimated, an intrinsic regime is,
contrary to expectations, not observed.\\[5mm]
\end{abstract}
%=======================================================================
\section{Introduction}

Under suitable conditions systems of statistical mechanics can exhibit
coexistence of different phases which are separated spatially by 
interfaces. Structure and properties of such interfaces have been
investigated both experimentally and theoretically, and still offer
problems for current research. Near critical points several interfacial
properties are expected to show universal behaviour.

Since the 19th century, there has been continuous interest in the
theoretical description of interfaces \cite{Wi72,RW82,Bi83,Ja84}.
Traditionally, the interface has been described by mean field theories
and their refinements \cite{vdW93,CH58,FW60} as an ``intrinsic''
continuous profile with a width proportional to the bulk correlation
length.

The existence of the interface, however, breaks the translational
invariance of the system, resulting in long-wavelength capillary wave
fluctuations of the interface position as Goldstone-modes. These
fluctuations are neglected by mean field theory which assumes a flat
interface, but they strongly influence the interfacial properties.
According to capillary wave theory \cite{BLS65,GF90}, the interface
width, for example, depends logarithmically on the system size and
diverges in the limit of an infinite system.

Mean field and capillary wave theory can be viewed as a small-scale and
large-scale description of the interface, respectively, and then be
combined in the ``convolution approximation'' \cite{Ja84,Wi72}. In this
picture the interface is described as an intrinsic profile which is
centered around a two-dimensional surface subject to capillary wave
fluctuations.

While capillary wave behaviour has been observed in many systems (e.g.\
experimentally on liquid-vapor interfaces \cite{expliquidvapor} and
polymer mixtures \cite{exppoly}, in simulations of polymer mixtures
\cite{WSMB9799} and of the Ising model \cite{BS83,MLS90,HP92}),
the mean field behaviour is difficult to access as the intrinsic
profile - if it has a well-defined meaning at all - is broadened by the
capillary wave fluctuations.

The concept of intrinsic versus large-scale interface structure has
so far not been defined unambiguously outside a given theory. There is
no clear ``snapshot'' definition of an interface, which always shows
overhangs, islands, clusters which may be as well assigned to the
phase-separating interface region as to the fluctuation structure of
the neighbouring homogeneous phases.

In this paper the attempt is made to separate the intrinsic structure
from the influence of the capillary waves via a blocking procedure
reminiscent of the Kadanoff block spin method of renormalization group
theory. The system of size $L\times L\times D$ with an interface
perpendicular to the $D$-direction is divided into blocks of size
$B\times B\times D$. The different length scales are separated by using
the block size $B$ as a variable cutoff. To calculate the interface
position of the strongly fluctuating interface profiles on small length
scales, we implement a method which respects translational invariance
(in contrast to various other approaches in the literature). In this
way, the local interface position, the interface profile and the
interface width can be studied on the different length scales, thereby
allowing a direct test of the convolution approximation and of the
limits of validity of the capillary wave theory and of the mean field
theory. It is found that on large length scales, the capillary wave
picture is consistent with the data. 

On small length scales, however, contrary to expectations we cannot
identify an intrinsic interface profile with a width of the order of
the correlation length. For the case of the Ising model under study the
concept of the intrinsic profile is thus called into question.

An approach based on blocking has already been used by Weeks \cite{We77}
for liquid-vapour interfaces in order to derive the capillary wave model
from a microscopic model Hamiltonian, thereby reconciling mean field
and capillary wave theory as small-scale and large-scale interface
descriptions, respectively. In the context of Monte Carlo simulations a
procedure similar to the one used in this paper has been applied by
Werner et al.\ \cite{WSMB9799} to homopolymer interfaces, finding
consistency of their data with the convolution approximation.

%=======================================================================
\section{Theoretical description of the interface\\ profile}
\label{theory}

\subsection{Intrinsic interface profile}

Interfaces can be described in mean field theory by a continuous profile
$\phi(x)$ representing the difference between the concentrations of the
two coexisting phases. In the framework of Landau theory and its field
theoretical extensions the profile function plays the role of a local
order parameter. Near the critical temperature the free energy density
is written as \cite{Be91}:
\begin{equation}\label{LandauFreeEnergyDensity}
  \mathcal{L} = \frac{1}{2}(\partial\phi(x))^2 + V(\phi(x))
\end{equation}
with the double-well potential
\begin{equation}
  V(\phi) = \lambda (\phi^2-1)^2\,.
\end{equation}
The minima of this potential correspond to the two homogeneous phases,
whose densities have been normalized to $\pm 1$ for simplicity.

Minimization of the free energy density (\ref{LandauFreeEnergyDensity})
with boundary conditions appropriate for an interface perpendicular to
the $z$-axis leads to the typical hyperbolic tangent mean field profile
\cite{CH58}
\begin{equation}
  \phi(z) = \tanh\left(\frac{1}{2\xi_{0}}(z-h)\right)
\end{equation}
with a width proportional to the mean field correlation length $\xi_0$,
which diverges near the critical point with exponent 1/2. The parameter
$h$ specifies the location of the interface which is arbitrary due to
translational invariance.

Corrections due to fluctuations of the order parameter field can be
calculated systematically in renormalized perturbation theory. In the
``local potential approximation'' only corrections to the local
potential $V(\phi)$ are taken into account, whereas contributions
involving higher powers of derivatives are neglected. The resulting
interface profile is of the form $\phi(z) = f((z-h) / 2\xi)$, where the
width is now proportional to the physical correlation length $\xi$,
which diverges near the critical point with exponent $\nu$.  This
corresponds to the scaling form proposed by Fisk and Widom \cite{FW60}.
To lowest order the profile function $f$ is equal to the mean field
profile function, $f=\tanh$. In higher orders, corrections to the
potential $V(\phi)$ due to perturbative loop contributions modify the
mean field profile \cite{KM04}. We do, however, not consider them as
they are numerically small and can be neglected in our context.

The interface profile
\begin{equation}\label{IntrinsicProfile}
  \phi(z) = \tanh\left(\frac{1}{2\xi}(z-h)\right)\,,
\end{equation}
in lowest order of the local potential approximation is to be
distinguished from the mean field profile. It represents a refinement of
Landau theory that is consistent with scaling \cite{FW60} and holds near
the critical point.

In the local potential approximation long wavelength fluctuations are
not fully taken into account. Consequently the profile
(\ref{IntrinsicProfile}) does not contain the effects of capillary
waves, which are discussed in the following section. The function above
thus represents an intrinsic interface profile.

%===================================
\subsection{Capillary wave theory}

Translational invariance is broken by the presence of the interface.
The Goldstone bosons associated with this broken symmetry are long
wavelength excitations of the interface position, which have vanishing
energy cost in the infinite wavelength limit \cite{GNW80}. These
capillary waves strongly influence interfacial properties like the
interface width but are neglected by mean field theory which assumes a
flat interface.  In capillary wave theory, as introduced by Buff
et al.~\cite{BLS65}, bubbles, overhangs and any
continuous density variations of the interface are neglected. The local
interface position is parameterized by a single-valued function
$h(x,y)$, describing the interface as a fluctuating membrane. The free
energy cost of capillary waves is basically due to the increase in
interface area against the reduced interface tension $\sigma$.  Thus the
capillary wave free energy (multiplied by the inverse
temperature $\beta$) is \cite{RW82}
\begin{equation}
  \beta \mathcal{H}
     = \sigma \!\int\!\!dx dy \left[\sqrt{1+(\nabla h)^2}-1\right]
     \approx \frac{\sigma}{2} \int\!\!dx dy\,(\nabla h)^2\,,
\end{equation}
where $|\nabla h| \ll 1$, i.e.\ long wavelength and small amplitude,
has been assumed. The Gaussian nature of the capillary wave model
allows analytic calculation of thermal averages. The local interface
position is distributed according to a Gaussian distribution
\begin{equation}\label{GaussianDistribution}
  P(h) = \langle\delta(h(x,y)-h)\rangle
       = \frac{1}{\sqrt{2\pi}\,s}\, \mathrm{e}^{-h^2/2s^2}
\end{equation}
with variance
\begin{equation}\label{variance}
  s^2
    = \langle h^2 \rangle = \frac{1}{2\pi\sigma} \int\!\!dq\,\frac{1}{q}
    = \frac{1}{2\pi\sigma} \ln \frac{L}{B_{\mathrm{cw}}}\,.
\end{equation}
In order to avoid divergence of the integral, a lower cutoff $2\pi/L$
and an upper cutoff $2\pi/B_{\mathrm{cw}}$ have been introduced. Here
$L$ is the system size, which determines the maximal allowed wavelength
and provides a natural lower cutoff for the capillary wave spectrum.
The cutoff length $B_{\mathrm{cw}}$, which is so far arbitrary, sets
the scale below which capillary wave theory ceases to be valid. As
there should be no capillary waves with wavelength smaller than the
intrinsic width of the interface, $B_{\mathrm{cw}}$ should be of the
order of the correlation length.  The factor $2\pi$ in the lower cutoff
is due to periodic boundary conditions and has also been introduced in
the upper cutoff for convenience. The phenomenological capillary wave
model thus requires two inputs: the macroscopic interface tension
$\sigma$ and the cutoff $B_{\mathrm{cw}}$. While the interface tension
near the critical temperature is known from scaling \cite{Ja84}, the
cutoff $B_{\mathrm{cw}}$ is unknown.

In the capillary wave picture the instantaneous interface profile is a
sharp step function between the two phases. Nevertheless, in the
thermal average the capillary wave fluctuations produce a continuous
density profile
\begin{equation}\label{CapillaryProfile}
  \rho(z)
     = \int\!\!dh\ \mathrm{sgn}(z-h)\,P(h)
     = \mathrm{erf} \left(\frac{z}{\sqrt{2}s}\right)
\end{equation}
with a finite width whose square is proportional to the variance $s^2$
(\ref{variance}) and thus to the logarithm of the system size $L$. In
other words, the apparent width of the interface depends on the length
scale on which the interface is studied and diverges in the
thermodynamic limit.

%===================================
\subsection{Convolution approximation}

The two different descriptions of the interface profile -  intrinsic
profile and capillary wave theory - can be reconciled by adopting the
view that the intrinsic profile describes the interface on a
microscopic scale of the order of the correlation length, while
capillary wave theory describes the macroscopic interface fluctuations
of wavelengths much larger than the correlation length. Neglecting the
coupling of the capillary wave fluctuations to the intrinsic interfacial
structure, the interface can be viewed as a fluctuating capillary wave
theory membrane $h(x,y)$, to which an intrinsic profile $\phi(z)$ is
``attached''. In this way an instantaneous profile $\phi(z-h(x,y))$ is
obtained. Performing the thermal average over the capillary waves, the
``apparent'' profile is given by the convolution
\begin{equation}\label{ConvolutionProfile}
  m(z) = \int\!\!dh\,\phi(z-h) P(h).
\end{equation}
In this convolution approximation, the intrinsic profile $\phi$ is thus
broadened by capillary wave fluctuations. The broadening increases with
the system size according to
Eqs.~(\ref{GaussianDistribution}) and (\ref{variance}) corresponding to
the fact that in a larger system more capillary waves are allowed.

%===================================
\subsection{Interface width}

The interface width cannot be defined in a unique way, and various
definitions have been used in the literature
(see e.g.\ refs.~\cite{Ja84}, \cite{BLS65} and \cite{MLS90}). For
numerical purposes, a suitable definition of the squared interface
width is the second moment of a weight function which is proportional
to the gradient of the interface profile m:
\begin{equation}\label{GradientWidth}
  w^2
  = \frac{\int\!\!dz\,z^2 m'(z)}{\int\!\!dz\,m'(z)}.
\end{equation}
Due to the linearity of this definition and of the convolution
approximation (\ref{ConvolutionProfile}), the squared width $w^2$ of
the convolution profile $m$ is simply the sum of the intrinsic width
squared $w_{\mathrm{int}}^2$ of the intrinsic profile $\phi$,
Eq.~(\ref{IntrinsicProfile}), and of the capillary width squared
$w_{\mathrm{cw}}^2 = s^2$ of the capillary wave profile $\rho$,
Eq.~(\ref{CapillaryProfile}):
\begin{equation}
  w^2_{\mathrm{ca}} = w_{\mathrm{int}}^2 + w_{\mathrm{cw}}^2
   = \frac{\pi^2}{3}\xi^2
   + \frac{1}{2\pi\sigma}\ln \frac{L}{B_{\mathrm{cw}}}.
\end{equation}
This expression shows the broadening of the intrinsic profiles by
capillary waves. It can be cast into a temperature-independent form by
expressing all lengths in units of twice the correlation length:
\begin{equation}\label{ConvolutionWidth}
  \hat{w}^2_{\mathrm{ca}} = \left(\frac{w}{2\xi}\right)^2
      = \hat{w}^2_{\mathrm{int}} + \hat{w}^2_{\mathrm{cw}}
      = \frac{\pi^2}{12}
       + \hat{A}_{\mathrm{cw}}
         \log_2 \frac{\hat{L}}{\hat{B}_{\mathrm{cw}}},
\end{equation}
where $\hat{L}=L/2\xi$, $\hat{B}_{\mathrm{cw}}=B_{\mathrm{cw}}/2\xi$.
Here the slope
\begin{equation}
  \hat{A}_{\mathrm{cw}}=\frac{\ln 2}{8\pi \sigma\xi^2}=0.265(2)
\end{equation}
contains the the universal constant $\sigma\xi^2 = 0.1040(8)$
\cite{HP97,HM98}. We also considered other definitions of the interface
width, e.g.\ with a weight function proportional to $(m'(z))^2$, but
the results to be discussed below are not altered significantly.

%=======================================================================
\section{Monte Carlo calculation}
\label{MC}

\subsection{Monte Carlo setup}

In experiments or Monte Carlo simulations of systems with interfaces
the observed interface profile and width are the total ones, including
the effects of the intrinsic structure as well as of the capillary
waves. In order to investigate the question whether the intrinsic
structure can be separated from the effects of capillary waves,
suitable observables have to be constructed. We have studied this
question by means of Monte Carlo simulations of the three-dimensional
Ising model, which is in the same universality class as
three-dimensional binary mixtures. The results allow to test the
predictions from capillary wave theory and the validity of the
convolution approximation.

In the simulations a cubic lattice of size $L\times L\times D$ is used,
with spins $s(x,y,z)\in\{\pm 1\}$ on each lattice site $(x,y,z)$. The
presence of an interface is enforced by antiperiodic
boundary-conditions in the $z$-direction, while the boundary conditions
in the $x,y$-directions are periodic. The lattice sizes are varied from
$L=32$ to 512 and from $D=50$ to 180. To reduce critical slowing down,
the Single-Cluster-Wolff algorithm \cite{Wo89} is used to allow
simulations at reduced temperatures
$t = (T-T_c)/T_c = -0.05$, $-0.01$ and $-0.004$, corresponding to
correlation lengths $\xi=1.65$, $4.56$ and $8.13$ in lattice units,
respectively.

In order to analyze the interface on different length scales a blocking
procedure is applied. The system is split into columns of block size
$B\times B$ and of length $D$. The block size $B$ will later be varied
to allow for systematic analysis. In each block $i$ a local block
profile is calculated by averaging over the lateral coordinates in the
block:
\begin{equation}\label{BlockProfile}
  m_i(z) = \frac{1}{B^2}\sum_{x,y\in\mathrm{Block}\ i} s(x,y,z)\,.
\end{equation}

%===================================
\subsection{Local interface position}

Due to strong thermal fluctuations near the critical point the
interface position for this profile is not well-defined. Various
methods have been proposed to determine the interface position
(see e.g.\ refs.~\cite{HP92}, \cite{HMP96} and \cite{St97})
and applied to total interface profiles, i.e.\ for the
profile (\ref{BlockProfile}) at $B=L$. 
These methods, however, are not appropriate for our purpose, because
they either neglect bulk fluctuations, whose contributions we want to
take into account fully, or they suffer from a lack of translational
invariance: If the interface is, due to
interface wandering, not in the middle of the system, the
fluctuations above and below the interface do not average out, leading
to a biased value for the interface position. This bias can be
neglected for profiles on sufficiently large length scales, i.e.\ for
block profiles with sufficiently large $B$, but not for the strongly
fluctuating profiles on smaller length scales $B$.

To solve this problem, a translationally invariant method to determine
the interface position is proposed \cite{Mueller}.\footnote{After
completion of this work ref.~\cite{SB92} has been pointed out to us,
where the same method is being used.} Because of the
pe\-ri\-odic/an\-ti\-pe\-ri\-odic boundary conditions, the Ising
lattice is wrapped to a torus. The antiperiodicity is implemented by
taking the couplings between nearest neighbours $s(x,y,z)$ and
$s(x,y,z+1)$ to be negative for a particular value of $z=z_a$.
Superficially, the choice of $z_a$ appears to break translation
invariance along the $z$-direction. This is, however, not the case.
Under a shift of $z_a$ the Hamiltonian remains invariant, if at the
same time those spins, which cross the plane of antiperiodicity, are
flipped. Consequently translation invariance remains a symmetry.
Observables should be constructed in such a way that they also respect
translation invariance. Translational invariance of our method to
determine the interface position is achieved in the following way. The
antiperiodicity point $z_a$ is shifted through the system successively.
The values of the block profiles $m_i(z)$, which cross the
antiperiodicity plane $z=z_a$, change signs appropriately. For a
monotone interface profile function $m(z)$, all values would get the
same sign, if $z_a$ coincides with the center of the interface.
Correspondingly, we define the location $z$ of the interface to be that
value of $z_a$, which makes the number of values of $m_i(z)$ having the
same sign maximal, i.e.\ being mostly positive or mostly negative.
Mathematically this point can be determined as the maximum of $|\sum_z
m_i(z)|$. In this way, one assures that the fluctuations near the
interface determine the precise location of the interface position. It
would also be possible to use the minimum of $|\sum_z m_i(z)|$, which
is attained when the interface is located opposite of $z_a$. In this
case, however, fluctuations deep in the bulk phase opposite (on the
torus) to the interface position would account for the precise location
of the interface position, which is physically less meaningful.

This boundary shift method very successfully determines the interface
positions for strongly fluctuating profiles even on rather small length
scales $B$, but of course fails for extremely small length scales
$B=1,2,3$, where the profiles $m_i(z)$ are still reminiscent of their
Ising nature with only two allowed values $\pm 1$ and thus do not
possess a well-defined interface position at all. This will result in
Ising lattice artefacts for extremely small block sizes in the later
discussion.  But as neither of the interface models discussed in
section \ref{theory} applies on a microscopic scale, this deficiency of
the method does not prevent a test of the models.

%===================================
\subsection{Interface profile}

In order to obtain the interface profile without fluctuations on length
scales larger than $B$, the block profiles $m_i(z)$ can now be shifted
such that the interface position is in the middle of the system. Note
that values that pass the antiperiodic boundary conditions change sign
(spin flip).  To assure that the profile values above the interface are
predominantly positive, an overall sign change is applied if necessary.
The shift procedure corresponds to a measurement in a moving frame that
is wandering with the local interface, i.e.\ to an elimination of the
zero mode within the block. One can now take the block and Monte Carlo
average over the block profiles $m_i(z)$ to obtain the local interface
profile $m(z)$ on the scale of the block length $B$. This profile
corresponds to the apparent profile (\ref{ConvolutionProfile}) of the
convolution approximation measured on a length scale of the block size
$B$.

The interface width of the profile is determined via a discretized
gradient analysis corresponding to Eq.~(\ref{GradientWidth}). Let the
$z$-coordinates take the values $1, 2,\ldots, D$. An auxiliary
coordinate $t=z-D/2$ is introduced which labels the positions between
adjacent lattice sites and is centered about the middle of the system,
assuming the values $-D/2+1, -D/2+1,\ldots, D/2-1$. A normalized
gradient is then defined by
\begin{equation}
  p(t) = \frac{1}{N} \left|m\left(t+\frac{D}{2}+1\right)
         - m\left(t+\frac{D}{2}\right)\right|
\end{equation}
with the normalization
\begin{equation}
N=\sum_t \left|m\left(t+\frac{D}{2}+1\right)
   - m\left(t+\frac{D}{2}\right)\right|\,.
\end{equation}
The interface width is the second moment of this weight function:
\begin{equation}
  w^2 = \sum_t t^2\,p(t) - \bigg(\sum_t t\,p(t)\bigg)^2.
\end{equation}

So far each block has been considered as an Ising system of size
$B\times B\times D$ in its own right, for which the same measurements
as for a total Ising system have been made. In this way all
fluctuations of wavelengths \textit{larger} than $B$ have been cut off.
The opposite approach is to first average over each block and then
consider the resulting Ising system of size $L/B\times L/B\times D$,
thereby ``averaging out'' all fluctuations of wavelength
\textit{smaller} than $B$. This coarse graining view is adopted in
order to obtain a capillary wave like description of the interface. In
each block $i$ the interface position $h_i$ is determined by the above
described boundary shift procedure and then measured relative to the
total interface position, i.e.\ the position of the profile $m_i$ for
$B=L$, determined by the same method.  The resulting height variables
for each Monte Carlo configuration form the snapshot of a membrane
corresponding to the function $h(x,y)$ of the capillary wave model, but
coarse grained on the length scale $B$.  According to the capillary
wave model, the distribution $P(h)$ of these height variables $h_i$
over all blocks and Monte Carlo configurations should be Gaussian,
Eq.~(\ref{GaussianDistribution}).

%=======================================================================
\section{Results}

\subsection{Confirmation of the capillary wave model}

\begin{figure}

\begin{minipage}{5.6cm}
\includegraphics[width=5.6cm]{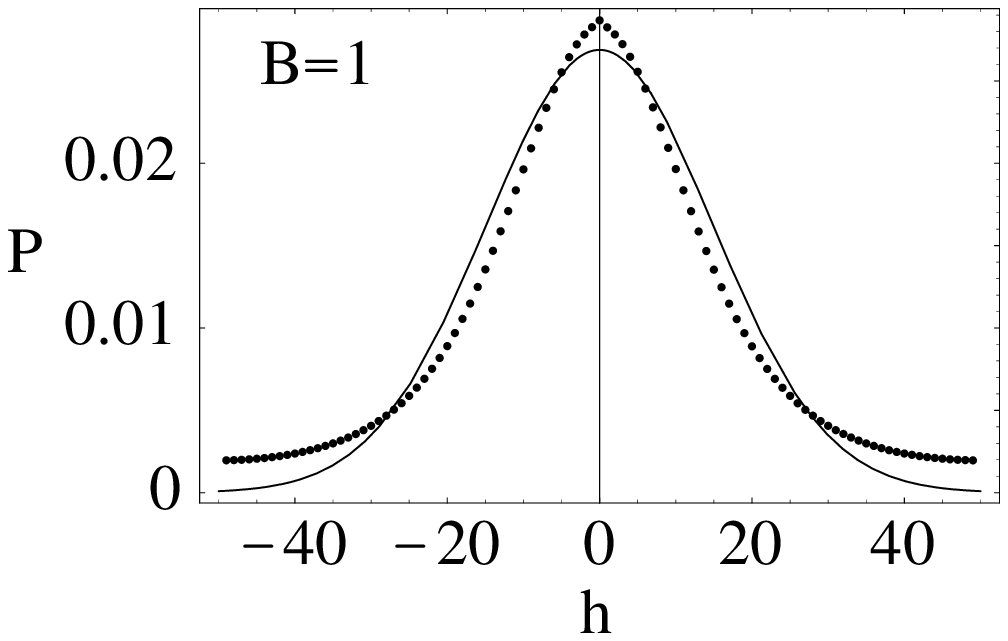}
\end{minipage}\hspace{0.1cm}
\begin{minipage}{5.6cm}
\includegraphics[width=5.6cm]{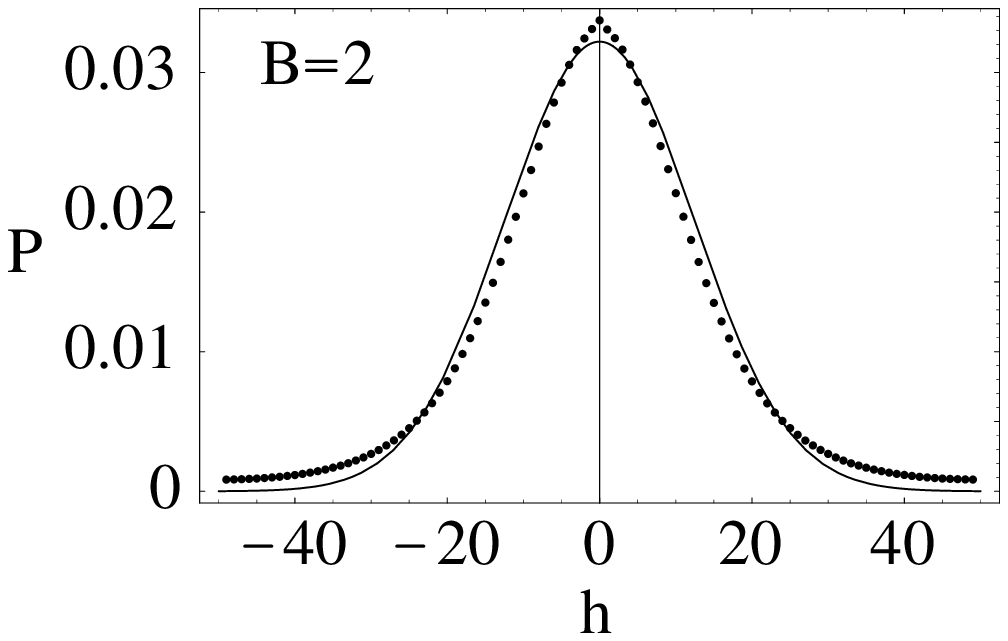}
\end{minipage}

\begin{minipage}{5.6cm}
\includegraphics[width=5.6cm]{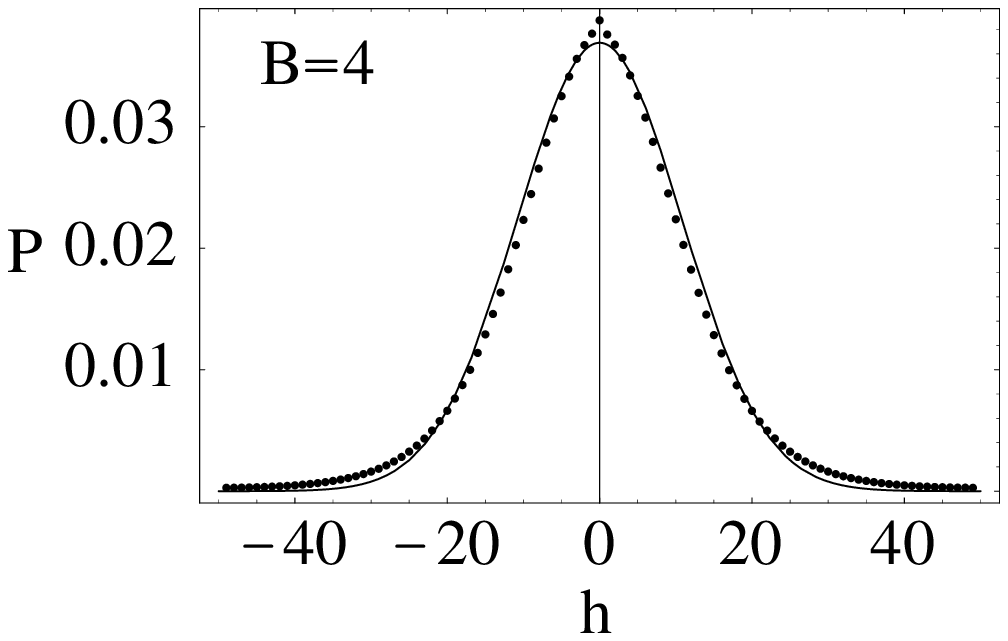}
\end{minipage}\hspace{0.1cm}
\begin{minipage}{5.6cm}
\includegraphics[width=5.6cm]{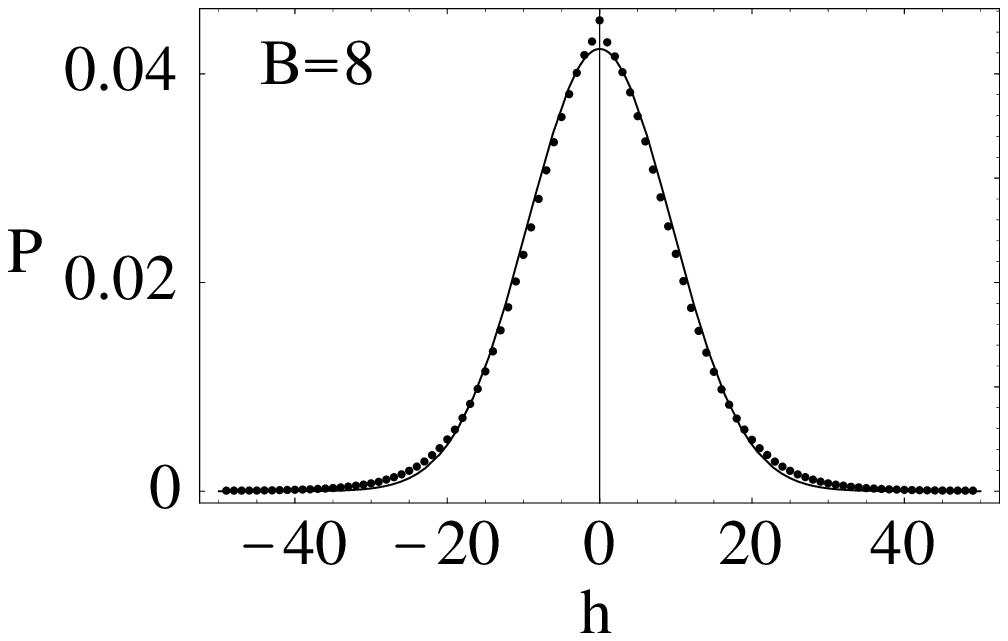}
\end{minipage}

\begin{minipage}{5.6cm}
\includegraphics[width=5.6cm]{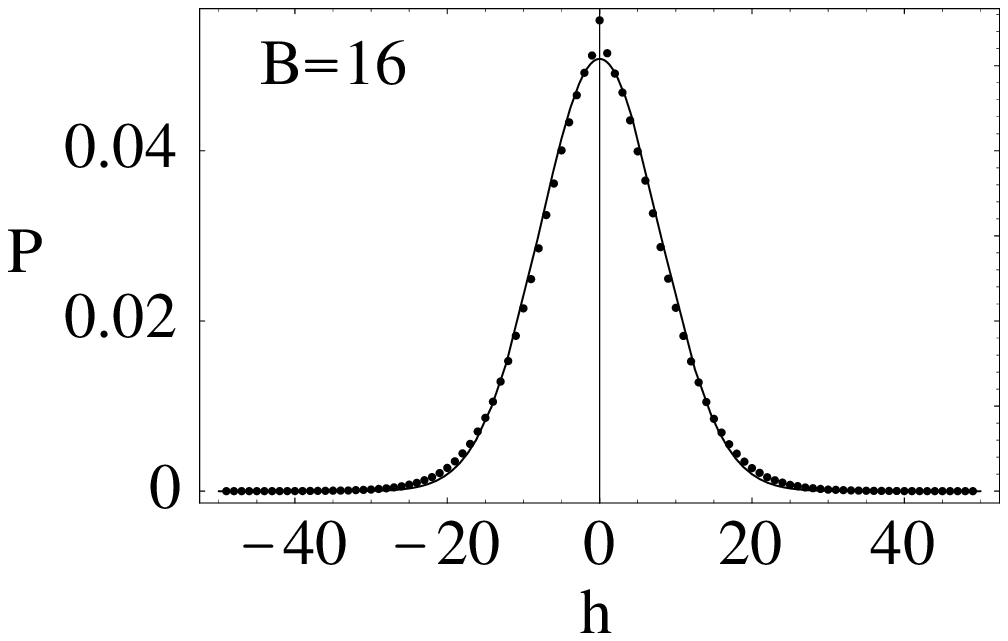}
\end{minipage}\hspace{0.1cm}
\begin{minipage}{5.6cm}
\includegraphics[width=5.6cm]{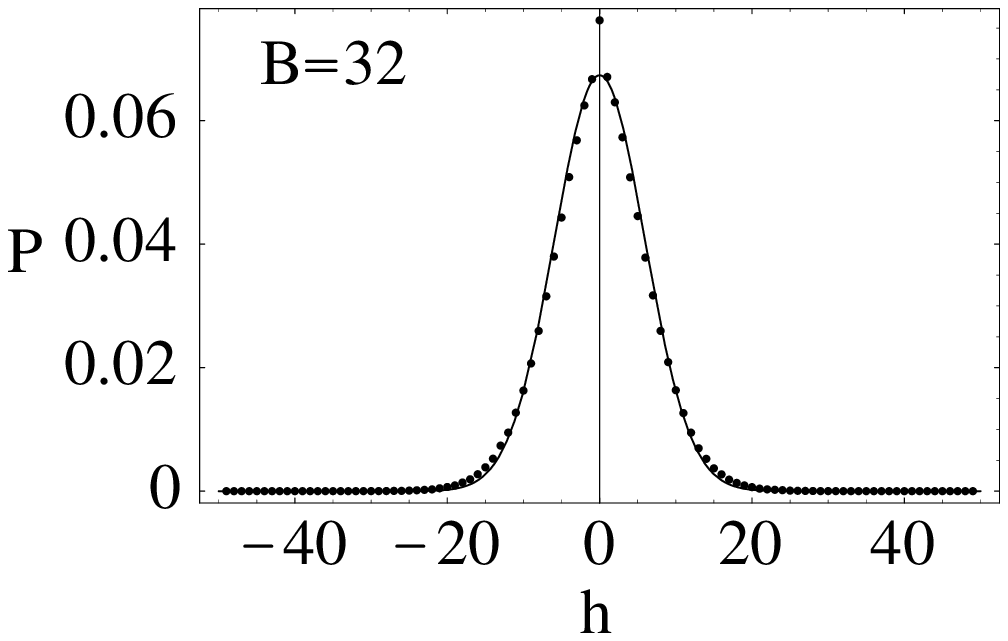}
\end{minipage}

\begin{minipage}{5.6cm}
\includegraphics[width=5.6cm]{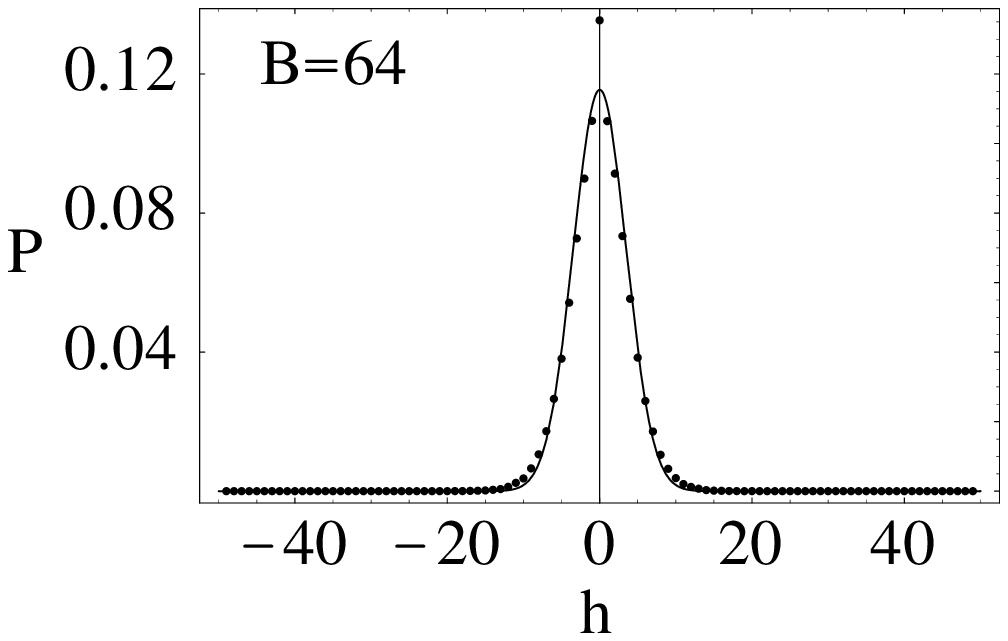}
\end{minipage}\hspace{0.1cm}
\begin{minipage}{5.6cm}
\includegraphics[width=5.6cm]{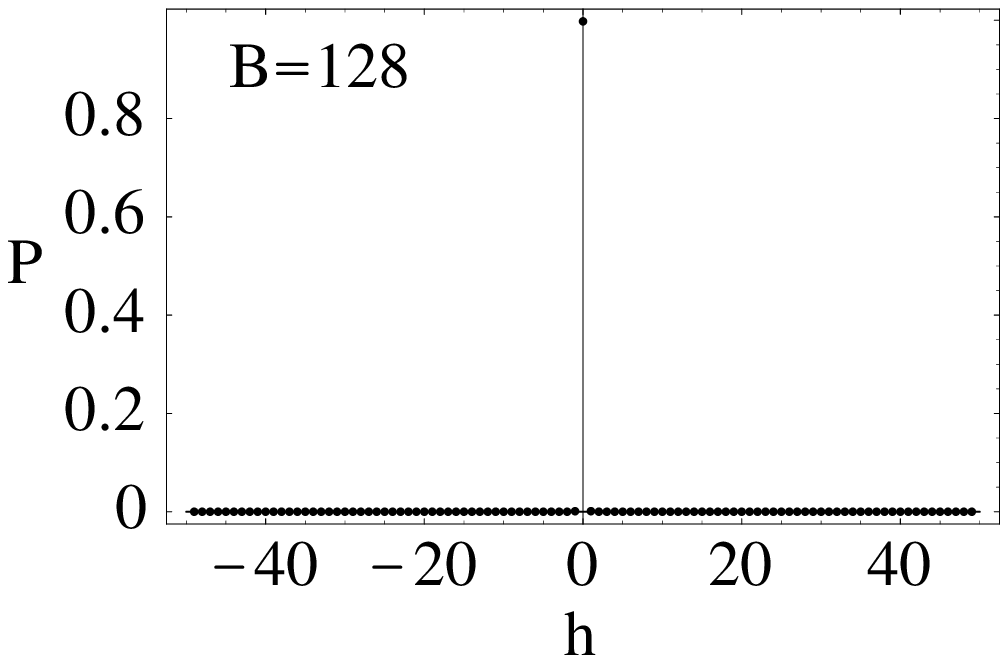}
\end{minipage}

\caption{Distributions $P(h)$ of the interface positions $h$, coarse
grained on different block sizes $B$, in a system of length $L=128$ at
reduced tempetature $t=-0.01$. In addition to the Monte Carlo data
(points), Gaussian fits are shown (lines).}
\label{FigDistributions}

\end{figure}

The measured distributions $P(h)$ of the interface heights at $t=-0.01$
and $L=128$ for various coarse graining lengths $B$ are shown in
Fig.~\ref{FigDistributions}.  They display Gaussian-like peaks which
become narrower with increasing block size, thus qualitatively
following the capillary wave model predictions
Eqs.~(\ref{GaussianDistribution}) and (\ref{variance}). For $B=L$ the
distribution has degenerated to a delta peak. The additionally
displayed Gaussian fits show that the measured peaks are leptocurtic
(i.e.\ are sharper and have longer tails than Gaussian distributions),
but become very close to a Gaussian distribution for larger block size.
This confirms the expectation that the macroscopic capillary wave model
is only valid on length scales larger than some cutoff
$B_{\mathrm{cw}}$ (compare the discussion after Eq.~(\ref{variance})).
For a quantitative analysis, a $\chi^2$-test has been performed for the
distributions $P(h)$ at temperatures $t=-0.05$ and $t=-0.01$, showing
consistency with a Gaussian distribution for block sizes $B$ larger
than
\begin{equation}\label{CutoffFromP}
  B_{\mathrm{cw}} \sim (5-7)\,\xi\,,
\end{equation}
which is a reasonable estimate of the capillary wave cutoff
$B_{\mathrm{cw}}$. Even for large block sizes, however, the value of
$P(h=0)$ is extraordinarily large. This is due to the lattice
discretization with too low resolution at the center of the Gaussian
peak.

The variances of the measured distributions $P(h)$ increase with the
logarithm of the system size and decrease with the logarithm of the
block size according to Eq.~(\ref{variance}). The prefactor
$1/(2\pi\sigma)$ is of the right order of magnitude, corresponding to
surface tensions $\sigma$ in the range between $0.033$ and $0.038$ for
$t=-0.05$. This agrees with the value $\sigma=0.035(1)$ obtained from
the scaling formula $\sigma=\sigma_{0} |t|^{2\nu}$ with
$\sigma_{0}=1.55(5)$ \cite{HP97} and $\nu=0.6304(13)$ \cite{GZ98}, and
with the value $\sigma=0.0342(2)$ from ref.~\cite{HP92}.

The capillary wave model thus yields an appropriate description of the
interface in the three-dimensional Ising model on length scales larger
than a couple of correlation lengths.

%===================================
\subsection{Test of the convolution approximation}

\begin{figure}
\includegraphics[width=9.5cm]{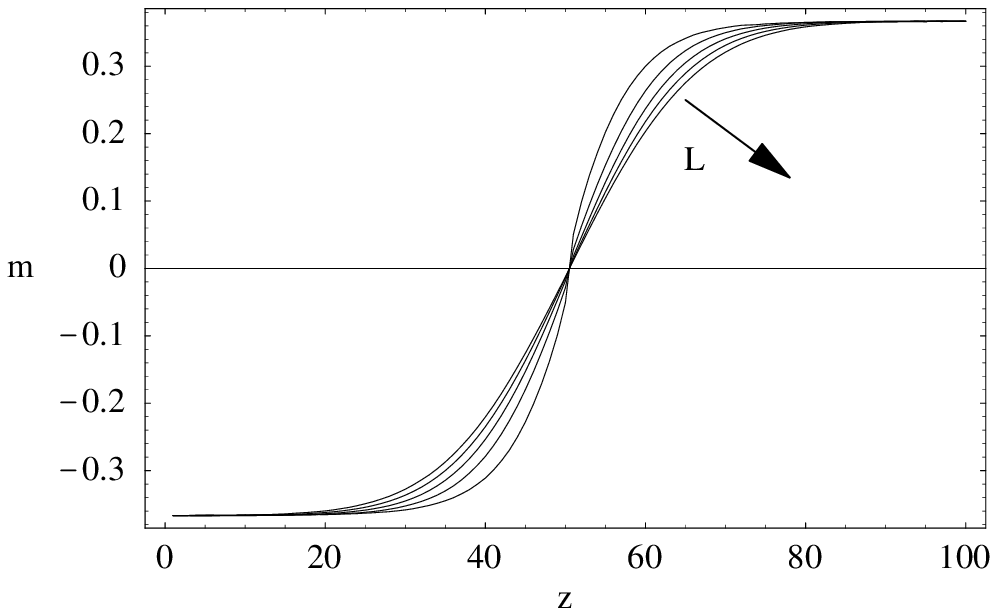}\\
\includegraphics[width=9.5cm]{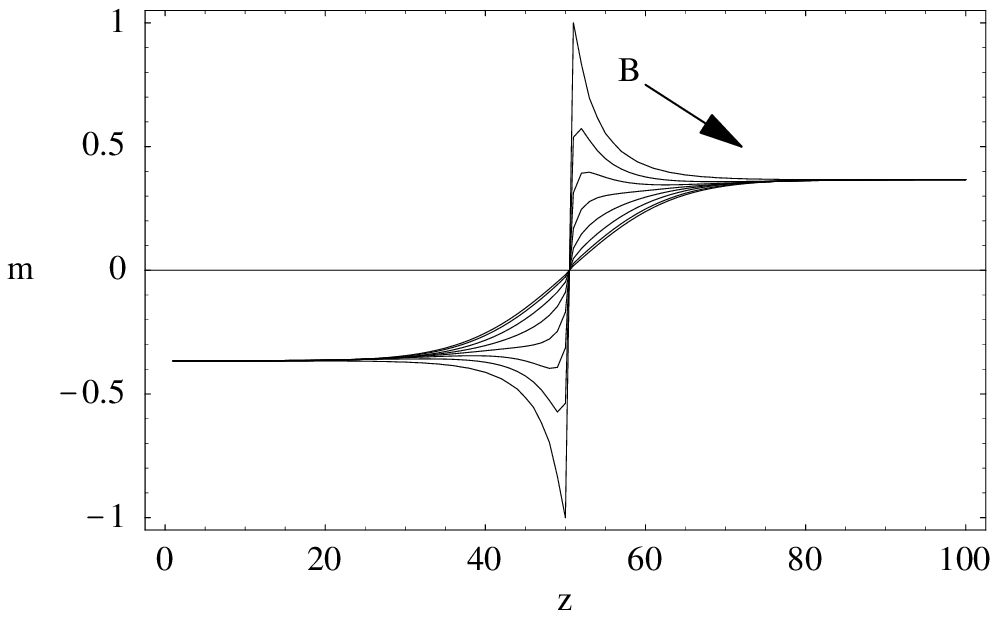}
\caption{%
Top: Total interface profiles (i.e.\ for $B=L$) for system lengths
$L=32,64,128,256$ and $512$.
Bottom: Interface profiles $m(z)$ for different coarse graining lengths
$B=1,2,4,\ldots,128=L$.
In both pictures the reduced temperature is $t=-0.01$.}
\label{FigProfiles}
\end{figure}

To probe the convolution approximation, Eq.~(\ref{ConvolutionProfile}),
the measured interface profiles $m(z)$ on various block sizes $B$ and
their widths $w^2$ are analyzed. The upper part of
Fig.~\ref{FigProfiles} shows the total interface profiles $m(z)$ (i.e.\
for $B=L$) for various system sizes $L$. The profile for $L=32$
displays a lattice artefact near the interface position, but the
profiles for larger system lengths are smoothly varying profiles of a
hyperbolic tangent or error function type as proposed by the
theoretical interface models of Section \ref{theory}. The broadening of
the profiles with increasing system size due to capillary waves is
clearly visible.

The lower part of Fig.~\ref{FigProfiles} shows the interface profiles
$m(z)$ on different block scales $B$. In the microscopic regime, i.e.\
for very small block sizes, lattice artefacts dominate and no
reasonable interface profile is obtained. With increasing block size,
however, the profiles approach a smooth hyperbolic tangent or error
function type profile which is then broadened with increasing block
size. This illustrates the renormalization group block spin procedure
leading from a microscopic Ising ``profile'' to the mean field profile,
which is then broadened by capillary waves, thus qualitatively
following the convolution approximation (\ref{ConvolutionProfile}).

\begin{figure}
\includegraphics[width=11cm]{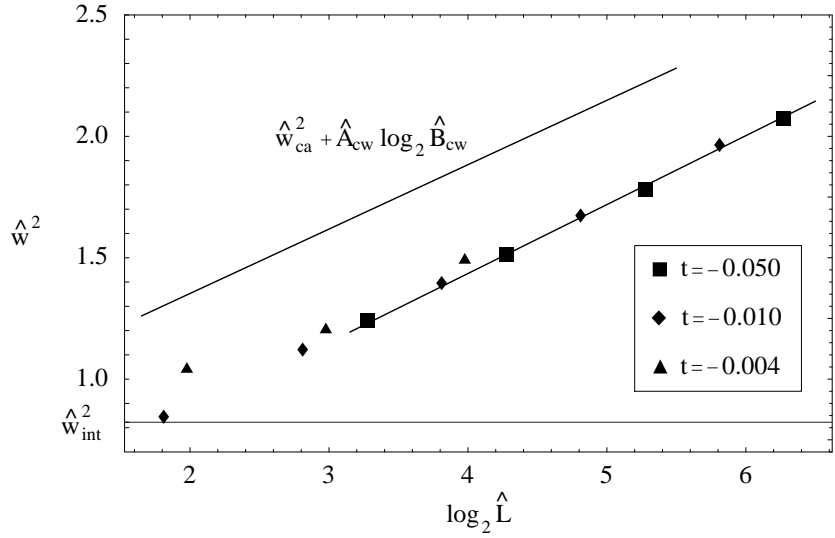}
\caption{Total interface widths $\hat{w}^2$ (in temperature-independent
units) of profiles for $B=L$ at the given temperatures $t$.
Statistical errors are smaller than the displayed symbols.
For comparison, the intrinsic width
$\hat{w}^2_{\mathrm{int}}=\pi^2/12$ and the prediction
$\hat{w}_{\mathrm{ca}}^2$ from the convolution approximation
(with an offset $\hat{A}_{\mathrm{cw}}\log_2\hat{B}_{\mathrm{cw}}$) are
shown.}
\label{FigWidthsEqualL}
\end{figure}

For a quantitative analysis the total interface widths 
$\hat{w}^2$ of the total interface profiles 
(i.e.\ for $B=L$) are plotted as functions of the logarithm of the 
system size $\hat{L}$ for various temperatures $t$, see
Fig.~\ref{FigWidthsEqualL}. As predicted by capillary wave theory and
scaling, Eq.~(\ref{ConvolutionWidth}), the width expressed in units 
of twice the correlation length all lie on the same
temperature-independent straight line, at least for large length
scales $\hat{L}$. The linear fit displayed in
Fig.~\ref{FigWidthsEqualL},
\begin{equation}\label{FitTotalWidth}
  \hat{w}^2 = 0.28(1)\log_2\hat{L} + 0.30(4)\,,
\end{equation}
is parallel to the theoretical prediction 
$\hat{w}^2_{\mathrm{ca}}$, Eq.~(\ref{ConvolutionWidth}), from
the convolution approximation with the slope
$\hat{A}_{\mathrm{cw}}=0.265(2)$.

The logarithmic divergence of the total interface width (for profiles
with $B=L$) in the Ising model has already been observed in other Ising
Monte Carlo simulations \cite{BS83,MLS90,HP92}, where the temperature
closest to the critical point was $t=-0.2$. Converting to
temperature-independent units, the obtained slopes are
$\hat{A}_{\mathrm{cw}}=0.222$ \cite{BS83}, $0.278$ \cite{MLS90} and
$0.244$ \cite{HP92}. The simulation by Stauffer \cite{St97} at $t=-0.01$
and $L=2000$ gave the single value $\hat{w}^2=0.99$, which is far below
our value $\hat{w}^2=2.49$ (extrapolated from
Eq.~(\ref{FitTotalWidth})). This discrepancy is due to Stauffer's
definition of the interface width which only measures the capillary wave
part of the width and eliminates the intrinsic contribution.

\begin{figure}
\includegraphics[width=11cm]{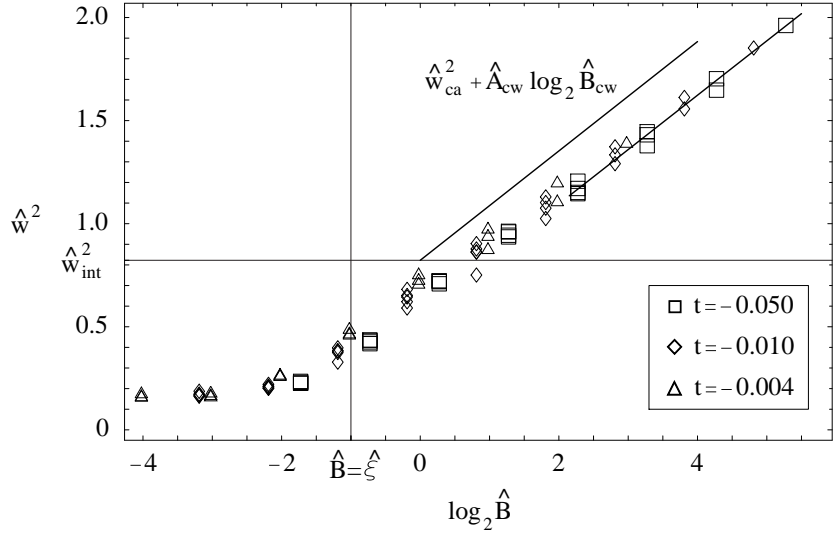}
\caption{Interface widths $\hat{w}^2$ (in temperature-independent units)
of profiles coarse grained on block lengths $B<L$.
Symbols are as in Fig.\ \ref{FigWidthsEqualL}. In addition, 
the correlation length $\hat{\xi}$ is shown.}
\label{FigWidthsSmallerL}
\end{figure}

Fig.~\ref{FigWidthsSmallerL} shows the widths $\hat{w}^2$ of the
profiles coarse grained on length scales $B<L$ as a function of the
logarithm of the block size $\hat{B}$. For each block size $\hat{B}$
the system shows a width $\hat{w}^2$ characteristic for this length
scale, independent of the temperature.  The variables expressed in
units of twice the correlation length thus indeed display
temperature-independent behaviour. Note that the widths of profiles on
scales $B<L$ have to be treated separately from those of total profiles
for $B=L$ due to different boundary conditions.  For $B=L$, periodic
boundary conditions apply, while for $B<L$ the boundary conditions are
dictated by the actual system configurations, resulting in a different
spectrum of capillary waves.

As expected from capillary wave theory, the widths grow with increasing
block size $\hat{B}$, and for large block sizes a linear increase with
$\log_2\hat{B}$ is obtained. The linear fit in
Fig.~\ref{FigWidthsSmallerL}
\begin{equation}
  \hat{w}^2 = 0.26(3)\log_2\hat{B} + 0.57(1)
\end{equation}
is parallel to the theoretical prediction $\hat{w}^2_{\mathrm{ca}}$,
Eq.~(\ref{ConvolutionWidth}), from the convolution approximation with
the slope $\hat{A}_{\mathrm{cw}}=0.265(2)$, thus again confirming the
capillary wave picture on large length scales.

In comparing the axis intercepts of the linear fits in
Figs.~\ref{FigWidthsEqualL}, \ref{FigWidthsSmallerL} to those of the
theoretical prediction (using as value for the intrinsic width
$\hat{w}^2_{\mathrm{int}}$ either the theoretical value
$\frac{\pi^2}{12}$, Eq.~(\ref{ConvolutionWidth}), or the ``microscopic''
value at small length scales as determined from
Fig.~\ref{FigWidthsSmallerL}, the parameter $B_{\mathrm{cw}}$ can be
determined. One obtains the reasonable values
\begin{equation}
  B_{\mathrm{cw}} \sim (3-7)\,\xi.
\end{equation}
of the same order of magnitude as in Eq.~(\ref{CutoffFromP}). The
cutoff $B_{\mathrm{cw}}$ can thus be consistently chosen to make theory
and simulation agree on large length scales.

%===================================
\subsection{Intrinsic width}

On smaller length scales the width in Fig.~\ref{FigWidthsSmallerL}
grows slower than linear with the block size and eventually becomes
saturated at a microscopic width $\hat{w}^2\approx 0.2$ for microscopic
length scales $B<\xi$. This behaviour is in qualitative agreement with
the convolution approximation. However, the microscopic width is
significantly smaller than the intrinsic width, which is expected to
have the much higher value $\hat{w}^2_{\mathrm{int}}=\pi^2/12$, see
Eq.~(\ref{ConvolutionWidth}).  The notion of an intrinsic profile
implies that the width should approach this value on the intrinsic
length scale, i.e.\ in some region around $B\sim\xi$.  In our
simulations such an intrinsic regime is not observed, contrary to
expectations.

The reason for the deviation from the expected picture is not clear.
The smallness of the interface width on the smallest scales suggests
that it is determined by Ising lattice artefacts. This could be
clarified by searching for the intrinsic profile in the Ising model
with larger correlation lengths, i.e.\ closer to the critical point, or
in $\phi^4$-theory.

%=======================================================================
\section{Conclusion}

In this article the interface in the three-dimensional Ising model has
been studied on different length scales. A procedure has been
implemented that allows to separate large and small wave length
contributions to the interface profile in a translation invariant way.
It is found that on large length scales the interfacial properties are
well described by capillary wave theory with a reasonable choice
$B_{\mathrm{cw}}\sim (3-7)\,\xi$ for the cutoff of the capillary wave
spectrum. The interface width shows universal behaviour, which is in
quantitative agreement with the theoretical predictions.  The data for
the widths are consistent with the convolution approximation which
includes both the intrinsic structure and the capillary waves. The
associated intrinsic width of the interface, however, turns out to be
much smaller than expected from field theory, and seems to represent a
``microscopic profile'' related to the discrete nature of the Ising
variables. An intrinsic profile could thus not be isolated for the
Ising model at the simulated temperatures.

Note added in proof: In ref.~\cite{CGMV} the width of colour flux tubes
in the 3-dimensional Z$_2$ gauge model has been investigated, which is
dual to the 3-dimensional Ising model. The results are in agreement
with ours.

%=======================================================================
%\newpage

%=======================================================================
%FIGURES%

%=======================================================================
\newpage

%=======================================================================
\newpage

\end{document}